\documentclass[12pt]{iopart}

\usepackage{graphicx}
\usepackage{hyperref}

\begin{document}

\title[Magnetic levitation on a type-I superconductor as an experiment for students]{Magnetic levitation on a type-I superconductor as a practical demonstration experiment for students}

\author{M R Osorio, D E Lahera and H Suderow}

\address{Laboratorio de Bajas Temperaturas, Departamento de F{\'i}sica de la Materia Condensada,
Instituto de Ciencia de Materiales Nicol{\'a}s Cabrera, Facultad de Ciencias, Universidad Aut{\'o}noma de Madrid,
E-28049 Madrid, Spain}
\ead{hermann.suderow@uam.es}
\begin{abstract}
We describe and discuss an experimental set-up which allows undergraduate and graduate students to view and study magnetic levitation on a type-I superconductor. The demonstration can be repeated many times using one readily available $25$ liter liquid helium dewar. We study the equilibrium position of a magnet that levitates over a lead bowl immersed in a liquid hand-held helium cryostat. We combine the measurement of the position of the magnet with simple analytical calculations. This provides a vivid visualization of magnetic levitation from the balance between pure flux expulsion and gravitation. The experiment contrasts and illustrates the case of magnetic levitation with high temperature type-II superconductors using liquid nitrogen, where levitation results from partial flux expulsion and vortex physics.
\end{abstract}

\submitto{\EJP}
\maketitle

\section{Introduction}
Meissner effect i.e., the expulsion of magnetic field from a superconducting material, has been extensively studied, both from the theoretical and experimental points of view \cite{Meissner1933,Brandt1989,Brandt1990,Chen1992,Strehlow2009,Badia-Majos2006,Schreiner2004,Valenzuela1999} \footnote{There are plenty of books where superconductivity phenomena are explained in depth, for instance in reference \cite{Tinkham}. A visual demonstration of superconductivity phenomena performed at the Michigan State University can be found in a set of $6$ videos. The address for the first one is given in reference \cite{MSU1965}. Meissner effect is described in segments $3$ and $4$.}.
To explain analytically the observed results different methods are used \cite{Saslow1991,Simon2001}, which are, for some simple situations, tractable by students. 
In type-I superconductors, like lead or aluminum, the Meissner effect provokes a total expulsion of the magnetic flux, and a magnet located on top of one of these materials is rejected due to repulsion. Nevertheless, typical demonstrations of magnetic levitation on a superconductor show a magnet levitating in apparent equilibrium on top of a superconductor. Of course, this occurs when using cuprate high $T_{\textrm c}$ superconductors, which are type II materials, where the magnetic field enters in the form of quantized vortices. When a magnet is brought close to a type-II superconductor, the behavior observed is the result of Meissner flux expulsion combined with vortex induced attraction or expulsion. Moreover, the magnetic history of the superconductor, with possible flux trapped, significantly influences the position of the magnet. Many related videos can be found in the web (see, for instance, references \cite{Dresden,UiO,UAM0,UAM01,UAM02}).

\begin{figure}
	\centering
		\includegraphics[width=0.5\textwidth]{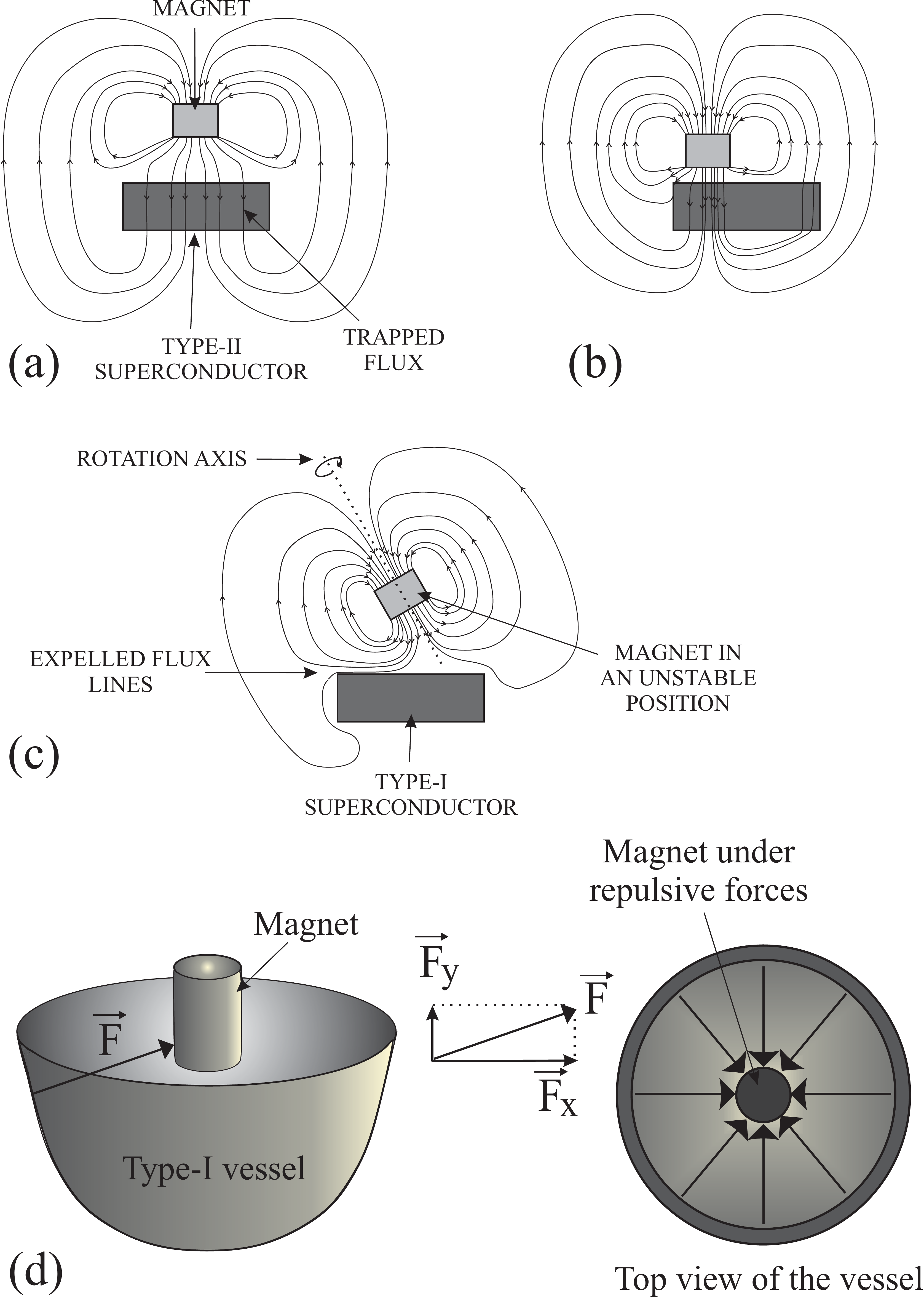}
	\caption{\footnotesize (a) Magnet levitating over a type-II superconductor. Pinned flux lines keep the magnet in a stable position. (b) Once the magnet is forced to move, another equilibrium position can be found. (c) Magnet over a type-I superconductor. As the magnetic flux is completely expelled from the superconductor, the magnet is not in a real stable position, and it may rotate or oscillate. (d) Magnet over a vessel made from a type-I superconductor. The horizontal components of the radial force allow a certain lateral stability, as depicted in the top view on the right.}\label{fig:Levitation-scheme}
\end{figure}

Vortices are pinned to imperfections in the sample (vacancies, impurities, dislocations, etc) and they tend to be arranged in a regular lattice (see, for instance, \cite{Essmann1967,Guillamon2009} and references therein). When the vertical position of the levitating magnet is altered just by pushing or unfastening, the vortex distribution is changed in the superconductor. This can occur as long as the pinning is not very strong. The same applies to lateral motion. In this way, in \Fref{fig:Levitation-scheme}(a) we can see a magnet located just above the center of the superconductor, with a symmetric distribution of vortices. In \Fref{fig:Levitation-scheme}(b) we push the magnet towards one side of the superconductor, and we obtain again an equilibrium position, although the vortex distribution has changed (we neglect demagnetizing effects due to superconductor's edges). On the contrary, when pinning is really strong the magnet can be difficult to move, and any small displacement is corrected as it tends to return to its initial position. Lateral stability can be taken for granted when using type-II superconductors, which is not the case for type-I, due to repulsion and the lack of flux penetration below the critical field, unless the intermediate state appears. This regime can arise in type-I superconductors, when the geometry of the superconducting body provokes that the density of flux lines is not homogeneous around its surface (for instance, for a sphere the density of flux lines would be higher at the equator and zero at the poles). Hence, the local magnetic field can be above the critical value, $H_{\textrm c}$, in some parts, whereas it remains below in other regions around the sample. This leads to the formation of normal and superconducting domains, whose frontiers are always parallel to the applied field, $H$. On the contrary, in the cross-section perpendicular to $H$, the distribution of normal areas exhibits peculiar and irregular domain patterns, depending on the geometry (sphere, foil, thin wire, etc). Note that the stable coexistence of these domains requires that the field in the normal regions is just $H_{\textrm c}$. Below this value the superconductivity would be regained, and above it the normal zones would propagate towards the superconducting adjacent regions \footnote{Further insight into this subject and nice images of the intermediate state can be found in \cite{Schmidt1997,Prozorov2008,Poole2007,Alers1957}.}.

In \Fref{fig:Levitation-scheme}(c) we can see a magnet levitating above the center of a type-I superconductor. Vortices do not exist because all the magnetic flux is expelled, and so there is pure repulsion. The lateral stability depends on the particular geometry of the system and the magnet can even exhibit a precession around a rotation axis \cite{Brandt1989,Brandt1990}.
This is avoided if the type-I superconductor is fabricated with a concave shape, like a bowl. The horizontal components of the radial repulsive forces are directed to the magnet in such a way that it can be stabilized around the center of the bowl \cite{Ouseph1990}. In \Fref{fig:Levitation-scheme}(d) we have depicted a scheme of a magnet levitating over a type-I superconducting vessel. The horizontal components of the repulsive force, $F_x$, keep the magnet in an equilibrium position at the center of the vessel. Note that the absence of a vortex lattice implies that there is only one equilibrium position.

In this work, we present a simple set-up to view magnetic levitation with a type-I superconductor. We propose an analytical procedure to calculate the equilibrium position of a cylindrical magnet that levitates over a vessel made from lead, whose critical temperature is $7.2$ K. We provide simple calculations, using basic magnetostatics, to explain the observed stable position of the magnet on top of the superconductor.

\section{Experimental procedure}\label{sec:Exp}

\begin{figure}
	\centering
		\includegraphics[width=0.5\textwidth]{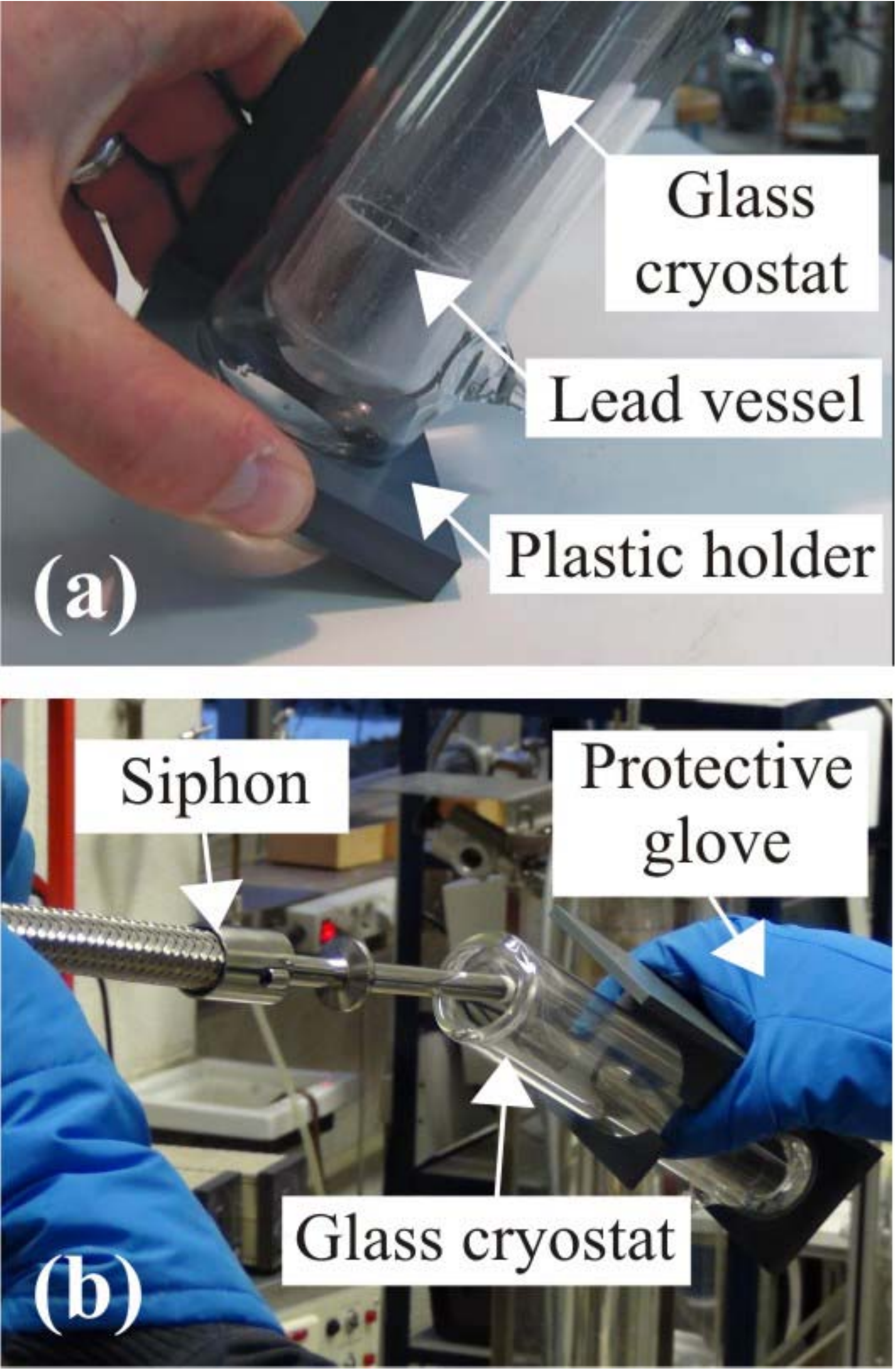}
\caption{\footnotesize (a) Photograph of the hand-held cryostat we used for the realization of the experiment.(b) The cryostat can be handled easily when pouring liquid helium, but using protective gloves is strongly recommended.}\label{fig:Dewar-Siphon}
\end{figure}

As mentioned above, the first step was the realization of the experiment with a vessel made from lead and a cylinder made from a NdFeB alloy (see the complete videos, recorded during the realization of the experiment, at \cite{UAM1,UAM2}). The vessel was inside a glass hand-held cryostat that has a double wall with vacuum insulation \footnote{More information about how to obtain small glass dewars where liquid helium is preserved for a few minutes can be obtained from authors. Made to measure glass tubes can be purchased at \cite{Dewar}.}. \Fref{fig:Dewar-Siphon}(a) shows a photograph of this dewar and the lead bowl.

As the critical temperature of lead is $7.2$ K, liquid helium is needed. Normally, this cryogenic liquid must be handled with care, and protective gloves and goggles must be worn. \Fref{fig:Dewar-Siphon}(b) shows how to hold the cryostat when pouring liquid helium. The siphon is of a common type and can be supplied with the helium bottle (safe handling and storage of liquid helium is briefly described in \cite{Safety}).
In addition, the latent heat of evaporation of helium-4 is much lower than that of liquid nitrogen ($20.6$ J/g and $199$ J/g, respectively, at their boiling temperatures \footnote{A comprehensive summary of the aspects involved in the handling of liquid helium can be found in the 5th chapter of reference \cite{White}.}) and so it evaporates very quickly  in a simple dewar as the one used here.

\begin{figure}
	\centering
		\includegraphics[width=0.5\textwidth]{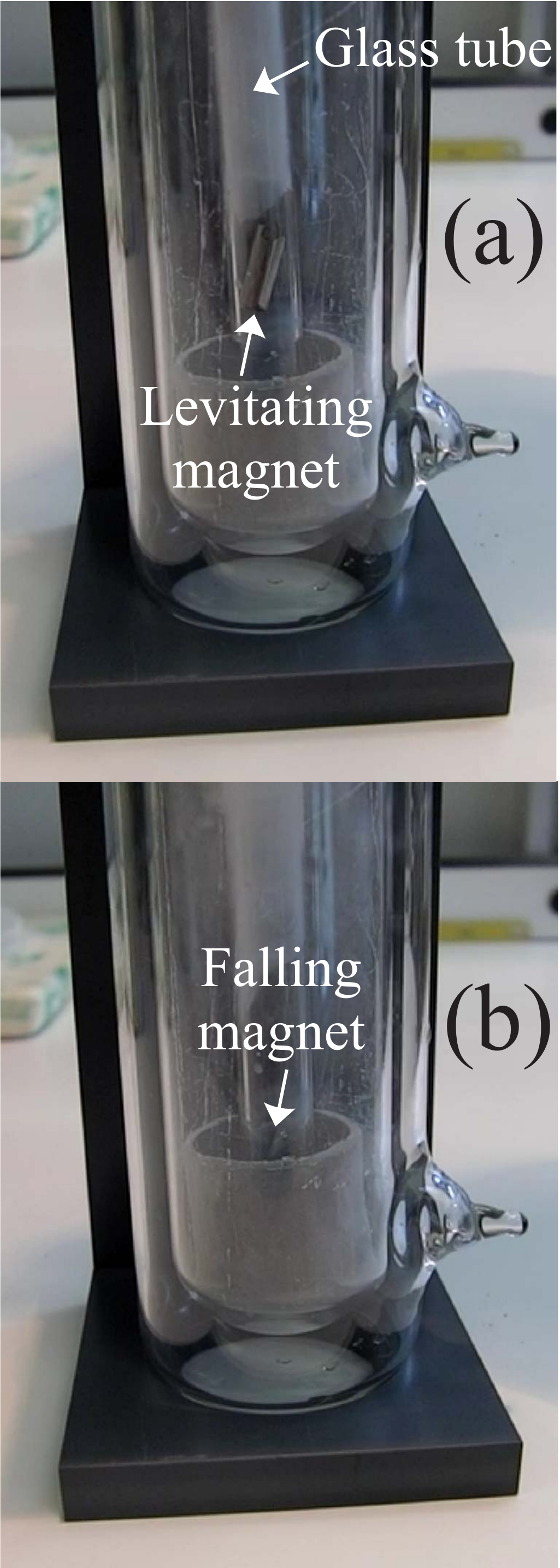}
	\caption{\footnotesize (a) Magnet levitating in an equilibrium position over the lead vessel. (b) Magnet falling down once the bowl leaves the superconducting state.}\label{fig:Magnet}
\end{figure}

We tried two different variations of this experiment. In both cases, the lead vessel was located in the bottom of the hand-held cryostat and a glass tube was used to guide the magnet towards the vessel. Nevertheless, some aspects were different, these being: (i) the magnet was let to slide down through the tube and towards the centre of the vessel after the cooling of the whole set-up. This is the experiment to be compared with our analytical model; (ii) a less powerful magnet was released through the tube but away from the symmetry axis of the vessel.  
Let us see these cases separately:

\begin{enumerate}
\item We first poured some liquid helium inside the cryostat. After some seconds, boil off relaxed and the lead bowl reached the full superconducting state. Hence, the glass tube was introduced to guide the magnet to the symmetry axis of the vessel and keep it in a vertical position. 
Once the magnet was let to slide inside the tube, it attained the lead bowl but, immediately, bounced and remained stable at an equilibrium position, about $1$ cm above the brim of the vessel (as can be seen in \Fref{fig:Magnet}(a), using as reference the height of the lead bowl, equal to $2.6$ cm).
As could be expected, displacing the glass tube did not alter the equilibrium distance. As noted above, the lack of pinning provoked that there was just one stable position, which could not be altered.
In the photographs it can observed that the magnet was slightly tilted, probably due to an inhomogeneous field distribution. This is not surprising, taking into account that the vessel was not totally regular, and hence the horizontal components of the repulsive forces could be somewhat unbalanced.  

After about $30$ seconds, once the liquid helium was completely evaporated, the lead transited to the normal state, and the magnet fell down the tube and inside the vessel, as shown in \Fref{fig:Magnet}(b).

\item In this experiment a small magnet was thrown in the cryostat, away from the symmetry axis of the vessel. In this case, it was completely repelled, and it was leaning against the cryostat walls even when this was rotated. This was the expected behaviour, taking into account the lack of symmetry of the forces when the magnet is apart from the vessel's centre.
\end{enumerate}

Concerning the refrigeration costs of these experiments, we can make a rough estimation of the liquid helium wasted during the whole run. First of all, the volume of liquid required to cool the magnet (the biggest one) from $300$ K to $4.2$ K can be calculated just by using its latent heat. Assuming that the Debye temperature of the magnet is $\Theta_{\textrm D}\approx 300$ K, and taking into account the Debye tables (see \cite{White}, section 11.2.5 and Table B2), the energy to be removed from the magnet is roughly $313$ J. As the latent heat of liquid helium is $2.6$ kJ/l, about $120$ ml of helium would be needed to cool the magnet.
Nevertheless, in a real case, a great deal of the cooling process is undertaken by the enthalpy of the gas (about $200$ kJ/l between $4.2$ K and $300$ K) and a lot of liquid is saved \cite{Pobell}. In fact, we estimate that we lose less than $30$ ml of liquid helium each time we cool down the vessel, the glass duct and the magnet.

\section{Calculation of the equilibrium position by using basic electrodynamics}
\subsection{Analytical model}

We present in this section a simple calculation to estimate the equilibrium position of the magnet on top of the superconductor. This calculation can be used in e.g., an electrodynamics course, together with the demonstration of the experiment.

\begin{figure}
	\centering
		\includegraphics[width=0.5\textwidth]{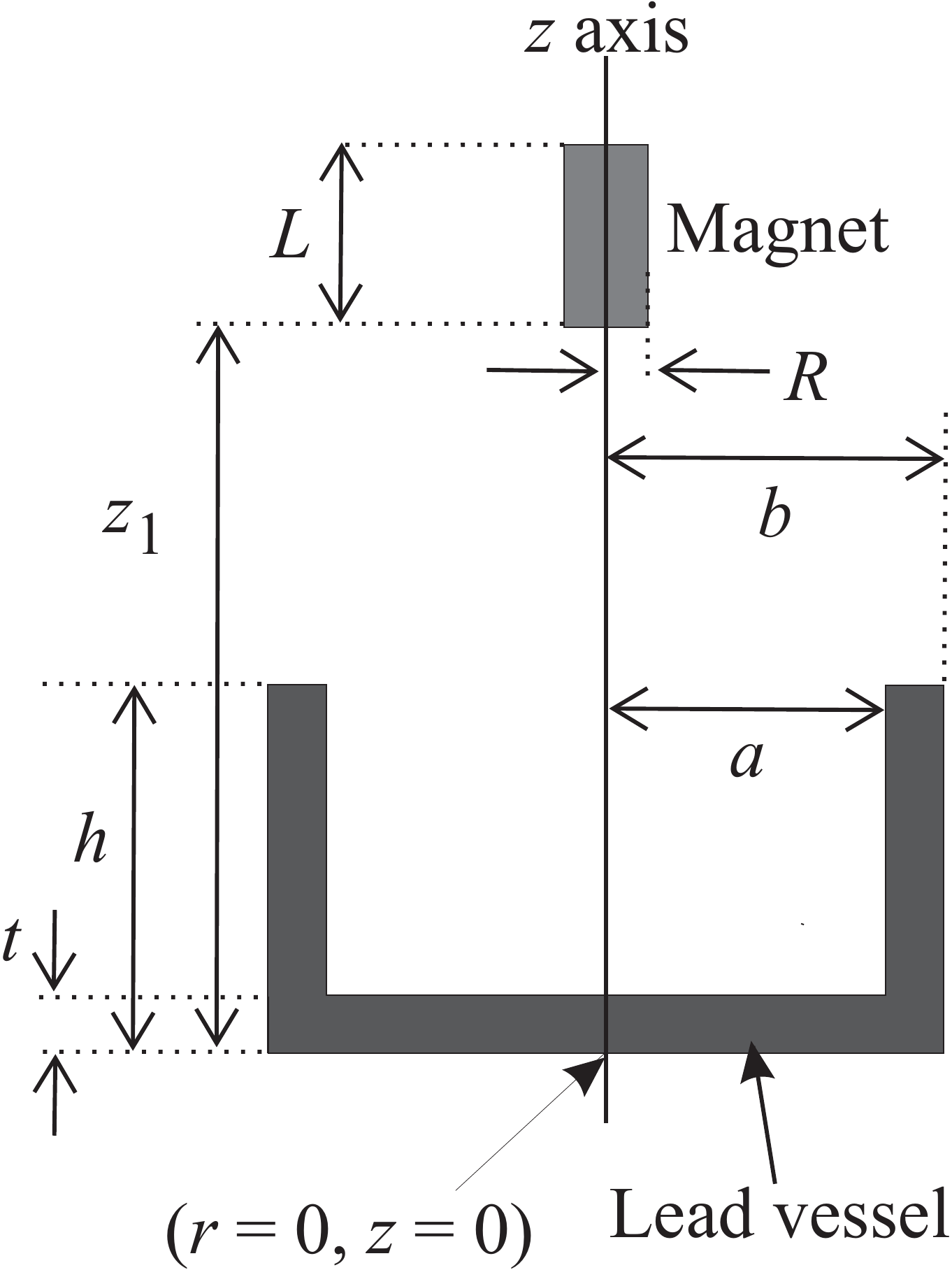}
	\caption{\footnotesize Cross-section of the system magnet-vessel and definition of the dimensions and distances which are going to be relevant in our calculations. The origin of the coordinate system is also indicated.}\label{fig:Scheme}
\end{figure}

As a first step, let us consider a cross-section of the system under study and a proper set of distances and dimensions, as depicted in \Fref{fig:Scheme}. In this scheme, $R$ and $L$ are the magnet's radius and length, respectively, $h$ the height of the lead vessel, $a$ its inner radius, $b$ the outer one, $t$ the thickness of its base and, finally, $z_1$ the distance between the bottom of the vessel and the magnet. The origin of coordinates is taken where the $z$ axis intersects the bottom of the vessel. The numerical values of all the parameters involved in the calculation of $z_1$ are gathered in \Tref{table:Values}.

\begin{table}[ht]
\caption{Dimensions of the magnet and the lead vessel.}
\centering 
\begin{tabular}{l c} 
\hline\hline 
Parameter and units& Numerical value \\ [0.5ex] 
\hline 
Magnet radius, R (mm) & 3  \\
Magnet length, L (mm) & 15\\
Magnet mass, m (g) & 3 \\
Vessel inner radius, a (mm) & 13  \\
Vessel outer radius, b (mm) & 15   \\ 
Vessel height, h (mm) & 26  \\
Vessel base thickness, t (mm) & 2   \\\\[1ex]  
\hline 
\end{tabular}
\label{table:Values}
\end{table}

\vspace{0.5cm}

In order to proceed with the calculations, we are going to consider a magnet having a constant magnetization oriented along the $z$ positive axis, i.e. ${\textbf M}=M{\bf \hat u_z}$ \footnote{To avoid carrying unnecessary subscripts, we are not going to use any special distinction for the magnitudes related to the magnet. The ambiguity is broken by the subscript ``vessel'' when referring to the lead bowl.}. 
The magnetization in the vessel can be calculated from the magnetic induction of the magnet as

\begin{equation}
{\textbf M_{\textrm {vessel}}}=-\frac{\textbf B}{\mu_0},
\label{Meissner}
\end{equation}
and it is related to the surface magnetizing current density, ${\textbf J_{\textrm {mS}}}$, through

\begin{equation}
{\textbf J_{\textrm {mS}}}={\textbf n} \times {\textbf M_{\textrm {vessel}}}.
\label{Currents}
\end{equation}

Note that in our superconducting vessel the thickness of this thin layer of current is the London penetration depth, and also that the constant magnetization precludes the existence of any volumetric current density i.e., $\textbf J_{\textrm {mV}}=\mathbf \nabla \times {\textbf M_{\textrm {vessel}}}=0$.

The repulsion force between the magnet (with constant induction $B$) and the vessel can be expressed as

\begin{equation}
{\textbf F_{\textrm m}}=\int_{S} {\textbf J_{\textrm {mS}}}\times{\textbf B}\, dA,
\label{Repulsion-Force}
\end{equation}
which, in turn, must balance the force exerted by gravity, so

\begin{equation}
{\textbf F_{\textrm m}}=-m_{\textrm {magnet}}{\textbf g}. 
\label{Weight}
\end{equation}

In order to solve \Eref{Repulsion-Force}, we need first to find a convenient expression for the magnetic induction.
Taking into account that there are no free currents inside the magnet, the magnetic induction can be expressed as the gradient of a magnetic potential i.e., ${\textbf B}=-\mu_0\mathbf \nabla\phi_{\textrm m}$.
Outside the magnet, this potential can be calculated by using \Eref{Phi} \cite{Jackson}:

\begin{equation}
\phi_{\textrm m}({\textbf r})= \frac{1}{4\pi}\int_{S}\frac{\sigma_{\textrm m}(\textbf r')}{\left | {\textbf r}-{\textbf r'} \right | }\, dA',
\label{Phi}
\end{equation}
where it has been used the magnetic pole surface density, defined as $\sigma_{\textrm m}({\textbf r'})={\textbf n}\cdot {\textbf M}({\textbf r'})$. Due to the magnetization orientation, $\sigma _{\textrm m}$ is just defined on both the top and bottom faces of the cylinder. Therefore, we will have $\sigma_{\textrm m}^{\textrm {top}}={\bf \hat u_z}\cdot {\textbf M}=M$ and $\sigma_{\textrm m}^{\textrm {bottom}}=-{\bf \hat u_z}\cdot {\textbf M}=-M$. We have not included the term corresponding to the magnetic pole volume density, $\rho_{\textrm m}({\textbf r'})=-\mathbf \nabla\cdot {\textbf M}({\textbf r'})$, which is zero because of the constant magnetization.
It is to be remarked that this approach, based on magnetic pole densities, is useful just outside the magnet, and that we would get erroneous results by applying \Eref{Phi} within its own volume.

Then, the scalar magnetic potential can be written as:

\begin{eqnarray}
\phi_{\textrm m}=\frac{M}{2}\int_{0}^{R} \frac{\rho'd\rho'}{\sqrt{(\rho-\rho')^2+(z_1+L-z)^2}}- \nonumber \\
\frac{M}{2}\int_{0}^{R} \frac{\rho'd\rho'}{\sqrt{(\rho-\rho')^2+(z_1-z)^2}},
\label{Potential-2}
\end{eqnarray}
where we have already performed the integration over the angular coordinate ($\int_{0}^{2\pi}d\varphi'=2\pi$). Note that the integral is calculated over the source radial coordinate, $\rho'$, which varies along the cylinder radius, whereas $z'$ is fixed and equal to $z_1$ for the bottom cylinder's base and $z_1+L$ for the top one.

By using a Taylor series and some convenient substitutions (see \ref{sec:a}), the scalar magnetic potential can be written in the following way:

\begin{eqnarray}
{\phi_{\textrm m}}\approx\frac{M}{4}\left(\frac{1}{(z_1-z)^3}-\frac{1}{(z_1+L-z)^3}\right) \nonumber \\
\left(\frac{R^4}{4}-\frac{2}{3}R^3\rho+\frac{R^2}{2}\rho^2\right)-\frac{M}{4}\frac{R^2L}{(z_1+L-z)(z_1-z)}.
\label{Potential-approx}
\end{eqnarray}

Now we can find out the magnetic field density by deriving this magnetic potential. As we are using a cylindrical coordinates system, the gradient components will be:

\begin{equation}
\mathbf \nabla\phi_{\textrm m} = \left(\frac{\partial\phi_{\textrm m}}{\partial \rho},\frac{1}{\rho}\frac{\partial\phi_{\textrm m}}{\partial\varphi},\frac{\partial\phi_{\textrm m}}{\partial z}\right).
\label{Gradient}
\end{equation}
The magnetic potential has not any angular dependence in this problem, so there are only two components of the magnetic field density, these being

\begin{equation}
B_\rho=-\mu_0\frac{\partial\phi_{\textrm m}}{\partial \rho},\;\;{\textrm {and}}\;\;B_z=-\mu_0\frac{\partial\phi_{\textrm m}}{\partial z}.
\label{Components}
\end{equation}

Applying the gradient function to \Eref{Potential-approx} yields\footnote{It must be remarked that this value of $B_\rho$ is valid just in the vicinity of the magnet, as commented in \ref{sec:a}.}:

\begin{eqnarray}
B_\rho\approx \mu_0\frac{M}{4}\left(\frac{1}{(z_1-z)^3}-\frac{1}{(z_1+L-z)^3}\right) \nonumber \\
\left(\frac{2}{3}R^3-R^2\rho\right).
\label{Br}
\end{eqnarray}
The other component, $B_z$, will be not calculated now because we will not use it to obtain the force in the ${\bf \hat u_z}$ direction, as we will see below.

Now that we have a suitable expression for $B$, we can calculate an analytical expression for the repulsion force. Combining \Eref{Currents} and \Eref{Repulsion-Force}, this force can be written as an integral over the whole vessel surface (see \ref{sec:b} for details):

\begin{equation}
{\textbf F_{\textrm m}}=\frac{1}{\mu_0}\int_{S} (B^2{\textbf n}-({\textbf B}\cdot{\textbf n}){\textbf B})\, dA.
\label{Repulsion-Force-2}
\end{equation}
This equation can be expressed as a sum of integrals corresponding to every face of the vessel. Then, taking into account the values of the magnetic field induction components, the repulsion force in the ${\bf \hat u_z}$ direction is (\ref{sec:b}):

\begin{eqnarray} 
{\textbf F_{\textrm m}}=-{\bf \hat u_z}\frac{2\pi}{\mu_0}\int_{0}^{b} \rho B_\rho^2(\rho,z=0)\,d\rho- \nonumber \\ 
{\bf \hat u_z}\frac{2\pi}{\mu_0}\int_{0}^{h} bB_\rho(\rho=b,z)B_z(\rho=b,z)\, dz+\nonumber\\ 
{\bf \hat u_z}\frac{2\pi}{\mu_0}\int_{a}^{b} \rho B_\rho^2(\rho,z=h)\, d\rho+ \nonumber \\
{\bf \hat u_z}\frac{2\pi}{\mu_0}\int_{t}^{h} aB_\rho(\rho=a,z)B_z(\rho=a,z)\, dz+\nonumber\\ 
{\bf \hat u_z}\frac{2\pi}{\mu_0}\int_{0}^{a} \rho B_\rho^2(\rho,z=t)\, d\rho. 
\label{Repulsion-Force-3}
\end{eqnarray}
The first member corresponds to the outer base, the second to the outer lateral face, the third to the brim, the fourth to the inner lateral face and the fifth to the inner base. Notice that we have made explicit the value for every non-integration variable. 

The analytical solution of \Eref{Repulsion-Force-3} does not yield a compact and easy-to-handle result, so we propose an approximation, this being that the magnetic field scarcely varies within the small thickness of the walls. In this way, it is assumed that the second and fourth terms in \Eref{Repulsion-Force-3} are virtually equal, except for their sign, and so its sum is close to zero, and the same applies to the first and the last terms. This is the reason why we do not need any analytical expression for the component $B_z$, as stated above (later on, we will analyse the quality of this approximation). Finally, only the third term of this equation survives, and we have a repulsion force given by (see \ref{sec:c}):

\begin{eqnarray}
{\textbf F_{\textrm m}}\approx \mu_0\frac{\pi}{8}M^2\mathcal{F}(R,a,b) \nonumber \\ 
\left(\frac{1}{(z_1-h)^3}-\frac{1}{(z_1´+L-h)^3}\right)^2,
\label{Repulsion-Force-4}
\end{eqnarray}
\newline
$\mathcal{F}(R,a,b)$ being a simple function of $R$, $a$ and $b$.

\subsection{Numerical result}

The magnetic induction created by the magnet is $B=4400$ Gauss \footnote{The characteristics and prices of these items can be found in \cite{Magnetization}.}, and so $M=B(\mu_0)^{-1}=0.35\;$MA/m.
Finally, by making \Eref{Repulsion-Force-4} equal to the force exerted by gravity (\Eref{Weight}), and by using the numerical values of \Tref{table:Values}, we get the following relation:

\begin{equation}
\left(\frac{1}{(z_1-0.026)^3}-\frac{1}{(z_1-0.011)^3}\right)^2=1.5\cdot 10^{12}.
\label{Numerical}
\end{equation}

Solving this equation by simple iteration yields a result of $z_1=3.52$ cm. Taking into account the definition of $z_1$ and $h$ given in \Fref{fig:Scheme}, we find that the distance between the brim of the vessel and the bottom of the magnet is $z_1-h=0.92$ cm. Despite the magnet did not levitate exactly at the center of the vessel and it was somewhat tilted, this result is in excellent agreement with the experimental value, as can be seen in \Fref{fig:Magnet}(a).

\subsection{Repulsion force calculated without approximations}

\begin{figure}
	\centering
		\includegraphics[width=0.5\textwidth]{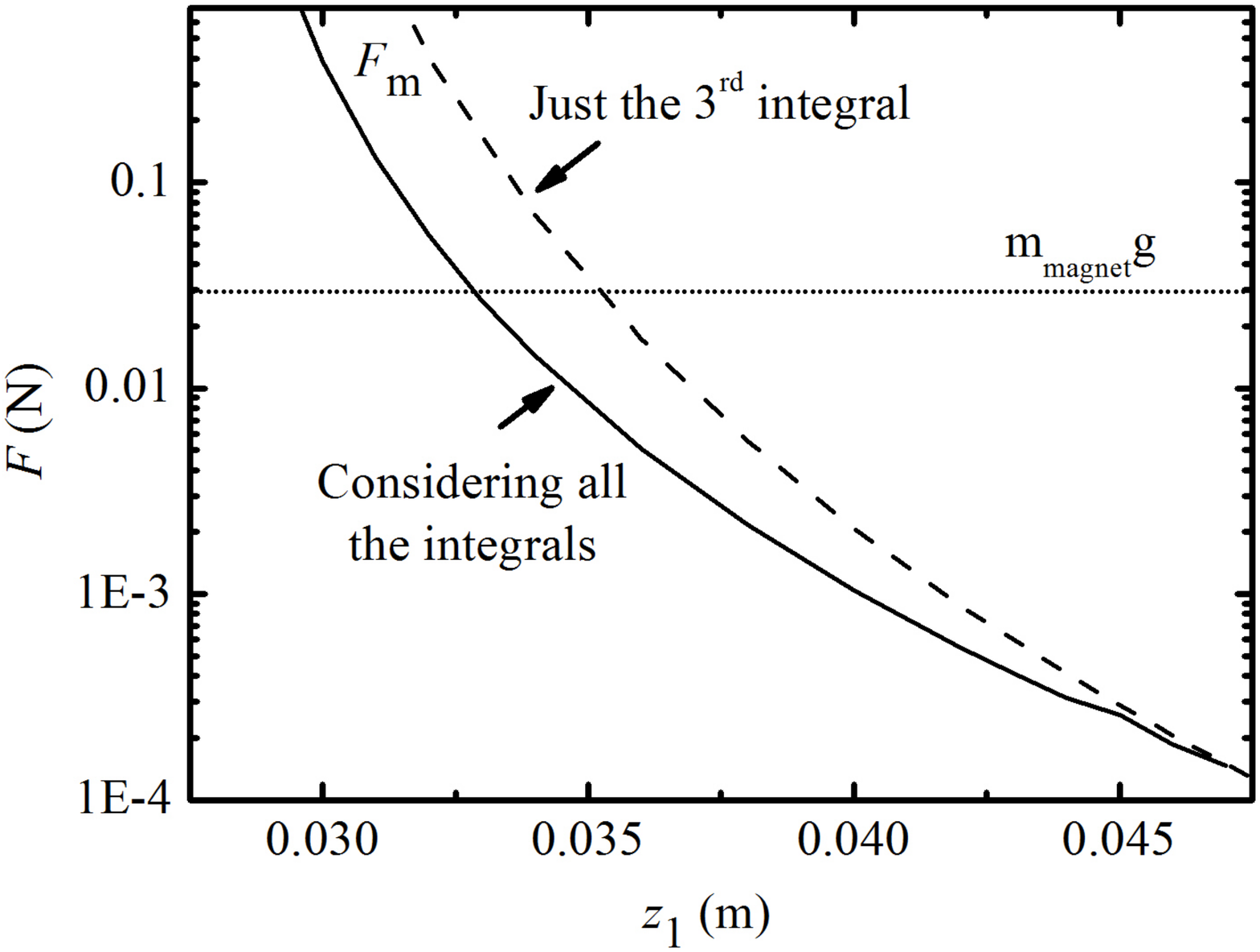}
\caption{\footnotesize Repulsion force calculated just with the third integral of \Eref{Repulsion-Force-3} (dashed line) and with the whole set of terms (solid line). They yield intersections with the gravity force (dotted line) which differ in about $2$ mm.} \label{fig:Numerical}
\end{figure}

For convenience, the exact solution of \Eref{Repulsion-Force-3} has been carried out by numerical integration. \Fref{fig:Numerical} shows the comparison of the force obtained just with the third integral (dashed line) and with the whole equation (solid line), as well as their intersections with the force exerted by gravity (dotted line) (see \ref{sec:d} for the analytical expression of $B_z$). It can be observed that the curves do not coincide, but there is a $2$ mm difference between both crossing points. Hence the new equilibrium distance would be $z_1-h=0.71$ cm.

The reason for the lack of accuracy of our approximation is that the integrals containing the component $B_z$ cannot be neglected (the first and fifth integrals do yield contributions which are several orders of magnitude smaller). Despite this error, the results are not so much different, and our calculation, intended to be as easy-to-handle as possible, allows obtaining quite a reasonable approach to the equilibrium distance.

\subsection{Considerations on the radial force}

Provided that the magnet is located along the symmetry axis of the vessel, the resultant of the radial force is zero (see appendix \ref{sec:b} to see how this is derived from calculations). This was already commented in the introduction and \Fref{fig:Levitation-scheme}(d), regarding the lateral stability of the magnet during the levitation. Nevertheless, if the magnet is out from the symmetry axis the scenario is different. In that case, there would be a non-zero resultant radial force, because of the unbalancing of the individual radial components, and the magnet would be repelled out from the vessel. This effect would explain the second case considered in section \ref{sec:Exp}, where a small magnet was expelled from the center of the vessel towards their borders.

Despite this, we observed that the cylindrical magnet we used for the experiment involved in the calculations was not repelled when its position was not exactly at the center of the vessel. This can be explained considering that it is a handmade bowl and not regular in shape, so the real equilibrium point for the magnet can be expected to be shifted to one side, or there may be different places where the magnet is more or less stable. In addition, the fact that the magnet is somewhat tilted with respect to the $z$ axis provokes changes in the balance of forces \cite{Perez-Diaz2009}.

On the other hand, it is possible that the off-axis equilibrium point is somehow related to the intermediate state, already mentioned in the introduction. This could be reasonable, taking into account that the cylindrical magnet is much stronger than the small magnet we used when we observed the total repulsion against the glass tube. In that scenario, some magnetic flux could have penetrated into the superconducting vessel, yielding a somewhat unpredictable equilibrium position.

\section{Conclusions}

We have reproduced and recorded in our laboratory the levitation of a magnet over a type-I superconductor by using a simple hand-held cryostat, and we have obtained from simple electromagnetism the analytical equation for the equilibrium position. Experiment provides for a visual discussion of the always fascinating levitation phenomenon. The pedagogical ingredients complement the popular levitation with type-II high Tc superconductors. Our experiment includes thermodynamics (properties and handling of cryogenic liquids) and tractable electrodynamics. Handling of liquid helium is greatly simplified with a hand-held small sized cryostat. The experiment can be carried out without a great cost in liquid helium. Using $25$ liters of liquid helium (approximately $250$ euros), we have shown the Meissner effect and discussed the properties of liquid helium to more than $200$ secondary school students each morning during $3$ days. This experiment also serves as an introduction to advanced solid state concepts (superconductivity), if combined with magnetic levitation in high $T_{\textrm c}$ materials. It can be used to discuss the applications of phenomena which are enabling industrial solutions of great importance to efficient and sustainable energy handling, such as cryogenics and superconductivity.

\appendix 
\section{Additional details on the calculations}

In this appendix we present some details on the different steps that are not strictly necessary to follow the calculation.

\subsection{}\label{sec:a}
\Eref{Potential-approx} was obtained by applying a Taylor series around $x=0$

\begin{equation}
\frac{1}{\sqrt{x+a}}=\frac{1}{\sqrt a}-\frac{x}{2a^{3/2}}+...,
\label{Ap-Taylor}
\end{equation}
and taking into account that $x=(\rho-\rho')^2$, and $a=(z_1+L-z)^2$ for the first term of \Eref{Potential-2} and $a=(z_1-z)^2$ for the second one\footnote{Note that this approximation implies that $\rho\to \rho'$ i.e, the field and source points are very close. Despite its apparent roughness, it yields good results for distances about the dimensions of our set-up.}.
In this way, we have \Eref{Ap-Phi}

\begin{eqnarray}
{\phi_{\textrm m}}\approx \frac{M}{4}\left(\frac{1}{(z_1-z)^3}-\frac{1}{(z_1+L-z)^3}\right)\int_{0}^{R} \rho'(\rho-\rho')^2\,d\rho' \nonumber \\
-\frac{M}{2}\frac{L}{(z_1+L-z)(z_1-z)}\int_{0}^{R} \rho'd\rho',
\label{Ap-Phi}
\end{eqnarray}
from which \Eref{Potential-approx} results after integration.

\subsection{}\label{sec:b}
\Eref{Repulsion-Force-2} was derived by substituting ${\textbf J_{\textrm mS}}=-\frac{1}{\mu_0}{\textbf n}\times {\textbf B}$ into \Eref{Repulsion-Force}, and then applying the vectorial identity $({\textbf a}\times{\textbf b})\times{\textbf c}=({\textbf c}\cdot{\textbf a}){\textbf b}-({\textbf c}\cdot{\textbf b}){\textbf a}$. Then, we expanded this last equation in terms of the different contributions due to each face of the vessel:

\begin{eqnarray}
{\textbf F_{\textrm m}}=\frac{1}{\mu_0}\int_{0}^{2\pi}\int_{0}^{b} (B^2(-{\bf \hat u_z})-({\textbf B}\cdot(-{\bf \hat u_z}){\textbf B}))\rho\,d\rho d\varphi+ \nonumber \\
\frac{1}{\mu_0}\int_{0}^{2\pi}\int_{0}^{h} (B^2{\bf \hat u_\rho}-({\textbf B}\cdot {\bf \hat u_\rho}){\textbf B})b\,dz d\varphi+ \nonumber \\
\frac{1}{\mu_0}\int_{0}^{2\pi}\int_{a}^{b} (B^2{\bf \hat u_z}-({\textbf B}\cdot{\bf \hat u_z}){\textbf B})\rho\,d\rho d\varphi+ \nonumber\\
\frac{1}{\mu_0}\int_{0}^{2\pi}\int_{t}^{h} (B^2(-{\bf \hat u_\rho})-({\textbf B}\cdot (-{\bf \hat u_\rho}){\textbf B}))a\,dz d\varphi+ \nonumber\\
\frac{1}{\mu_0}\int_{0}^{2\pi}\int_{0}^{a} (B^2{\bf \hat u_z}-({\textbf B}\cdot {\bf \hat u_z}){\textbf B})\rho\,d\rho d\varphi. 
\label{Ap-Repulsion-Force}
\end{eqnarray}

In addition, it was taken into account that $B^2=B_{\rho}^2+B_z^2$, ${\textbf B}\cdot {\bf \hat u_{\rho}}=B_{\rho}$, ${\textbf B}\cdot {\bf \hat u_z}=B_z$, as well as $\int_{0}^{2\pi}{\bf \hat u_z}d\varphi=2\pi{\bf \hat u_z}$, and $\int_{0}^{2\pi}{\bf \hat u_\rho}d\varphi=0$ (let us remember that ${\bf \hat u_\rho}=\cos\varphi{\bf \hat u_x}+\sin\varphi{\bf \hat u_y}$, which yields zero when integrated between $0$ and $2\pi$).

\subsection{}\label{sec:c}
\Eref{Repulsion-Force-4} was obtained when substituting the calculated expression for $B_\rho$ into the third term of \Eref{Repulsion-Force-3}, as commented, yielding

\begin{eqnarray}
{\textbf F_{\textrm m}}\approx \mu_0\frac{\pi}{8}M^2\left(\frac{1}{(z_1-h)^3}-\frac{1}{(z_1+L-h)^3}\right)^2 \nonumber \\
\int_{a}^{b} \rho\left(\frac{2}{3}R^3-R^2\rho\right)^2\,d\rho. \nonumber \\
\label{Ap-Repulsion-Force-4}
\end{eqnarray}
After integration, it was obtained \Eref{Ap-Repulsion-Force-5} 

\begin{eqnarray}
{\textbf F_{\textrm m}}\approx \mu_0\frac{\pi}{8}M^2\left(\frac{1}{(z_1-h)^3}-\frac{1}{(z_1+L-h)^3}\right)^2 \nonumber \\
\left(\frac{2}{9}R^6(b^2-a^2)-\frac{4}{9}R^5(b^3-a^3)+\frac{1}{4}R^4(b^4-a^4)\right). \nonumber \\
\label{Ap-Repulsion-Force-5}
\end{eqnarray}
Notice that the term involving powers of $R$, $a$ and $b$, is the function $\mathcal{F}(R,a,b)$ of \Eref{Repulsion-Force-4}.

\subsection{}\label{sec:d}

The component $B_z$, resulting from direct derivation of \Eref{Potential-approx} with respect to $z$, is

\begin{eqnarray}
B_z\approx -\mu_0\frac{3M}{4}\left[\left(\frac{R^4}{4}-\frac{2R^3}{3}\rho+\frac{R^2}{2}\rho^2\right)\left(\frac{1}{(z_1-z)^4}-\frac{1}{(z_1+L-z)^4}\right)\right] \nonumber \\
+\mu_0\frac{M}{4}LR^2\left[\frac{1}{(z_1-z)(z_1+L-z)^2}+\frac{1}{(z_1-z)^2(z_1+L-z)}\right]
\label{Bz}
\end{eqnarray}

Although calculating the second and fourth integrals of \Eref{Repulsion-Force-3} is not difficult, the process is tedious and the resulting analytical equation is very long and not easy to handle. For that, as a first approximation, only the third integral was used to obtain the repulsion force.

\section{Acknowledgments}
The Laboratorio de Bajas Temperaturas is associated with the ICMM of the CSIC. This work
was supported by the Spanish MICINN (Consolider Ingenio Molecular Nanoscience CSD2007-
00010 program and FIS2008-00454 and by ACI-2009-0905), by the Comunidad de Madrid
through the program Nanobiomagnet (S2009/MAT1726) and by the NES program of the ESF.

\end{document}